\documentclass[12pt]{iopart}

\usepackage{graphicx}
\usepackage{amssymb}
\usepackage{amsfonts}
\usepackage{bm,bbold}

\begin{document}

\title[Newton's cradle with BECs]
      {Newton's cradle analogue with Bose-Einstein condensates}

\author{Roberto Franzosi$^{1,2}$ and Ruggero Vaia$^{2,3}$}

\address{$^1$ QSTAR and Istituto Nazionale di Ottica,
             Consiglio Nazionale delle Ricerche,\\
             largo Enrico Fermi 2,
             I-50125 Firenze, Italy}
\address{$^2$ Istituto Nazionale di Fisica Nucleare, Sezione di Firenze,
             via G.~Sansone 1, I-50019 Sesto Fiorentino (FI), Italy}

\address{$^3$ Istituto dei Sistemi Complessi,
             Consiglio Nazionale delle Ricerche,
             via Madonna del Piano 10,
             I-50019 Sesto Fiorentino (FI), Italy}
\ead{roberto.franzosi@ino.it}

\begin{abstract}
We propose a possible experimental realization of a quantum analogue
of Newton's cradle using a configuration which starts from a
Bose-Einstein condensate. The system consists of atoms with two
internal states trapped in a one dimensional tube with a longitudinal
optical lattice and maintained in a strong Tonks-Girardeau regime at
maximal filling. In each site the wave function is a superposition of
the two atomic states and a disturbance of the wave function
propagates along the chain in analogy with the propagation of
momentum in the classical Newton's cradle. The quantum travelling
signal is generally deteriorated by dispersion, which is large for a
uniform chain and is known to be zero for a suitably engineered
chain, but the latter is hardly realizable in practice. Starting from
these opposite situations we show how the coherent behaviour can be
enhanced with minimal experimental effort.
\end{abstract}



\submitto{\jpb}

\maketitle


\section{Introduction}
\label{s.intro}

Classical  machines represent a smart way to transmit insight of
physical mechanisms concealed into nature. Quantum Mechanics has been
one of the most prolific sources of unexpected phenomena, but is
often hard to understand. Thus, finding a classical machine which is
a paradigm for the quantum nature of a system is an engrossing
challenge. Often, when one succeeds in fulfilling such a task, a
plethora of phenomena intrinsic to the quantum nature can be observed
and exploited to highlight the difference between the quantum and the
classical realm.

In this work we trace a route towards the possible experimental
realization of a quantum analogue of the Newton's cradle (NC) system:
the analogy requires~ ({\em{i}}) a one-dimensional array of~
({\em{ii}}) individual quantum systems, representing the spheres in
the NC, and~ ({\em{iii}}) a nearest-neighbour interaction between
them, modelling the contacts between the spheres. This last point is
more evident if one thinks of a classical NC with spheres not in
close contact but slightly separated: the momentum is transmitted
between neighboring spheres in a short finite time and travels along
the array towards the last sphere. Note that the beautiful experiment
reported in~\cite{weiss} under the title {\em A quantum Newton's
cradle} does not meet the above points, because what is observed
there are the opposite-phase oscillations of two macroscopically
populated coherent states created from a Bose-Einstein condensate
within a single quasi-harmonic potential well. In other words the
paper~\cite{weiss} is a fascinating demonstration of how coherent
states are the quantum analogue of classical particles.

In order to meet the above requirements for realizing a quantum NC
(QNC), we consider a collection of cold atoms trapped in a
one-dimensional periodic potential, which can be built by confining a
Bose-Einstein condensate in a one-dimensional tube that constrains it
to a strict Tonks-Girardeau regime, whose first experimental
realization has been reached in the remarkable experiment
of~\cite{paredes}, with a set-up closely similar to the one
considered here. Superimposing a further optical potential of
moderate amplitude along the longitudinal direction generates an
optical lattice fulfilling condition ({\em{i}}). The dynamics of this
system is effectively described by a one dimensional Bose-Hubbard
model and, in the Tonks-Girardeau regime, the strong repulsive
interaction between the atoms prevents the double occupancy of
lattice sites. For the condensate we consider atoms with two possible
internal states, say $|0\rangle$ and $|1\rangle$, like the hyperfine
states of Rubidium atoms used in~\cite{MandelGWRHB2003} to create
entangled states; in this way each potential well hosts an effective
two-state system ({\em{ii}}) and the wave-function at each lattice
site is given by a superposition of these internal states. The
tunnelling interaction between nearby wells, which can be globally
tuned by the intensity of the optical lattice beam, provides the
required coupling, which meets condition ({\em{iii}}).

\begin{figure}[t]
\hfill\includegraphics[width=130mm,clip=true]{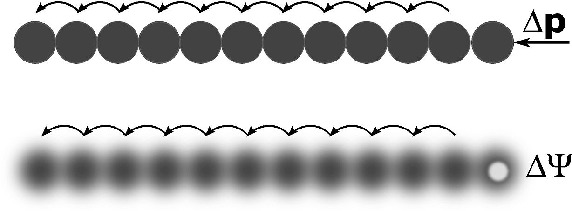}
\caption{The analogy between the classical Newton's cradle, where the
impulse of the mechanical momentum is transmitted along nearby
spheres, and the realization through a Bose -Einstein condensate,
where a wave-function disturbance is `delivered' along the optical
lattice.}
\label{analogy}
\end{figure}

In the following it is shown that a local perturbation generated at
one end of the lattice propagates back and forth between the lattice
ends in a very similar way to that in which an initial momentum pulse
is periodically exchanged between the endpoint spheres of the
classical NC: in fact, the role of the classical momentum
$\Delta{\bm{p}}$ transferred between the chain ends is played now by
the wave-function disturbance $\Delta\Psi$ which is transmitted
through the system. In particular, we assume the lattice prepared
with all sites in (say) the $|0\rangle$ state, and the initial
disturbance $\Delta\Psi$ consists in changing the first site to the
$|1\rangle$ state: the latter will propagate through the `sea' of
$|0\rangle$ states. The analogy is exemplified in
figure~\ref{analogy}.

The system and its dynamics are introduced in Section~\ref{s.system};
in Section~\ref{s.uniform-perfect} we consider two antithetical
cases: the simplest uniform lattice, with an unluckily poor dynamics,
and the perfectly transmitting lattice, with individually constructed
couplings. Note that in both these cases the net number-of-atoms
transport is null, while a counterflow
dynamics~\cite{prlKuklovSvistunov} is active. In fact, the currents
of the two atomic internal states have the same intensity and
opposite signs. Thus, in the case of the uniform lattice the
counterflow is dissipative, and in the second setup a
super-counter-fluidity is observed~\cite{prlKuklovSvistunov}. These
two examples help to understand under which conditions signals are
transported efficiently between the lattice ends and why in some
cases they fade out. As realizing the second setup in the lab would
be a strong challenge, we examine in Section~\ref{s.qperfect} two
arrangements that involve slight adjustments to the least demanding
uniform lattice and yield very high dynamic quality. The results are
summarized and discussed in Section~\ref{s.conclusions}.


\section{The system}
\label{s.system}

We consider a system of atoms with two internal states subjected to a
strong transverse trapping potential with frequency
$\omega_\perp\gg\mu/\hbar$, where $\mu$ is the chemical potential,
with a further standing-wave laser beam applied so that a periodic
potential in the longitudinal direction is created. For sufficiently
strong transversal and longitudinal potential, and low temperatures,
the atoms will be confined to the lowest Bloch band. The low-energy
Hamiltonian is then given (see~\ref{a.BHmodel}) by the Bose-Hubbard
model for two boson
species~\cite{prlKuklovSvistunov,ahdlNJP03,SoylerCPS2009} labeled by
$\alpha\,{=}\,0,1$:
\begin{eqnarray}
 H &=& \sum_{\alpha = 0}^1
 \sum_{j=1}^M \big[ U_\alpha \hat n_{\alpha j}(\hat n_{\alpha j}-1)
 + \xi_j \hat n_{\alpha j}\big]
\nonumber\\
 && +U \sum_{j=1}^M (\hat n_{0 j}-1/2) (\hat n_{1 j}-1/2)
\nonumber\\
 && -\sum_{\alpha=0}^1  \sum_{j=1}^{M-1} t_{\alpha{j}}\,
 \big(\hat{a}^{\dagger}_{\alpha j} \hat{a}_{\alpha,j+1}
 + \textrm{h.c.} \big) \, ,
\label{BHH}
\end{eqnarray}
where the index $j$ runs on the lattice sites and the boson operator
$\hat{a}_{\alpha j}$ ($\hat{a}_{\alpha j}^\dagger$) destroys
(creates) an atom in the internal state $\alpha$ at the lattice site
$j$. Boson commutation relations $[\hat{a}_{\alpha
j},\hat{a}^{\dagger}_{\beta j'}] =\delta_{jj'}\delta_{\alpha\beta}$
are satisfied and $\hat{n}_{\alpha j}=\hat{a}^{\dagger}_{\alpha
j}\hat{a}_{\alpha j}$ is the boson number operator for the species
$\alpha$ at site $j$. In the expression above, $U$ and $U_\alpha$ are
the inter- and intra-species on-site interaction energy,
respectively, whereas $t_{\alpha j}$ is the amplitude for the species
$\alpha$ to hop between sites $j$ and $j{+}1$. It is realistic to
assume that the site energy offsets $\xi_j$ do not depend on the
internal atomic state.


\subsection{Effective dynamics in the Tonks-Girardeau regime}
\label{ss.effdyn}

The homogeneous one-dimensional Bose-Hubbard model, has two
remarkable dichotomous limits. In the case of a vanishing repulsion
($U=U_\alpha=0$), the model described by Hamiltonian~\eref{BHH},
reproduces two independent ideal Bose gases on a lattice. The case we
are interested in is instead the opposite one, that is with a strong
repulsive interaction ($U,U_\alpha\to\infty$) and with a number of
atoms equal to the number of sites. In the latter case, an ideal
Fermi gas is found. As a matter of fact, very high values of
$U/t_{\alpha j}$ and $U_\alpha/t_{\alpha j}$, for $j=1,\dots ,M-1$,
entail such a high amount of energy for accumulating more than one
atom in a given site, that no site can be doubly occupied. Hence,
assuming $U,U^\alpha\gg{t}_{\alpha j}$, the only observable states
are those where the occupancy of any site is equal to one:
$\hat{n}_{0j}{+}\hat{n}_{1j}=1$. These states form a restricted
Hilbert subspace ${\cal{H}}_{\rm{R}}$. A first consequence of this
constraint is that the offset term in~\eref{BHH}, $\sum_j\xi_j$, is a
constant that we may neglect.

The two possible one-atom states at site $j$ can be written as
$|0\rangle_j\equiv\hat a^\dagger_{0j}|00\rangle_j=|10\rangle_j$, and
$|1\rangle_j\equiv\hat a^\dagger_{1j}|00\rangle_j=|01\rangle_j$,
where $|00\rangle_j$ is the empty state for the $j$-th site: hence,
$|0\rangle_j$ and $|1\rangle_j$ correspond to the $j$th atom in the
internal state 0 or 1, respectively. In this way the dynamics is
ruled by the only internal states and an effective Pauli exclusion is
realized. More precisely, noting that double creation or annihilation
are prohibited, i.e., $\hat{a}_{\alpha
j}^\dagger\hat{a}_{\alpha'j}^\dagger=0 =\hat{a}_{\alpha
j}\hat{a}_{\alpha'j}$, we consider the two operators
$(\hat{a}_{0j}^\dagger\hat{a}_{1j})$ and its conjugate
$(\hat{a}_{1j}^\dagger\hat{a}_{0j})$. They are easily shown to
satisfy Fermi anticommutation relations; this is true on the {\em
same} site $j$: operators on {\em different} sites are still
commuting. A full Fermi algebra can be recovered by means of the
Jordan-Wigner transformation
\begin{equation}
 \hat{c}_j=\exp{\textstyle\Big(\rmi\pi\sum_{\ell=1}^{j-1}\hat{n}_\ell\Big)}
           ~\hat{a}_{0j}^\dagger\hat{a}_{1j}~,
\label{e.JW}
\end{equation}
where the Fermion number
\begin{equation}
 \hat n_j\equiv\hat{c}_j^\dagger\hat{c}_j=\hat n_{1j}=1{-}\hat n_{0j}
\end{equation}
takes values 1 and 0 that match the atom species populating site $j$.
The Fermionic vacuum $|0\rangle_j$ corresponds to a 0-atom, while
$|1\rangle_j$ is now interpreted as the one-Fermion state. Therefore,
an overall generic state in ${\cal{H}}_{\rm{R}}$ takes the form of a
superposition of $2^M$ base states,
\begin{equation}
 |s\rangle = \sum_{\{\alpha_1,\dots,\alpha_{_M}\}}
 \Gamma_{\alpha_1,\dots,\alpha_{_M}}
 |\alpha_1\dots\alpha_{_M}\rangle ~.
 \label{st}
\end{equation}

The dynamics of the system in the Hilbert subspace
${\cal{H}}_{\rm{R}}$ follows from~\eref{BHH} and can be described by
an effective Hamiltonian $H_{\rm{R}}$ which can be found by the
method used in Refs.~\cite{prlKuklovSvistunov,praPachosRico}. The
general $H_{\rm{R}}$ is obtained in~\ref{a.fermioni}, while we
reasonably assume here that in~\eref{BHH} the interaction constants
do not depend on the atomic internal state, i.e., $U_0=U_1=U$ and
$t_{0j}=t_{1j}\equiv{t_j}$, which eventually yields the following
\begin{equation}
 H_{\rm{R}} = - \sum_{j=1}^{M-1} \tau_j\, (\hat c^\dagger_{j} \hat c_{j+1}
           + \hat c^\dagger_{j+1} \hat c_{j}) ~,
\label{Hfermi}
\end{equation}
with $\tau_j\,{=}\,t_j^2/U$: indeed, as the expectation values of the
hopping term in the Hamiltonian~\eref{BHH} vanish within
${\cal{H}}_{\rm{R}}$, the possible dynamical processes occur at
second order; in other words, to preserve site occupation two
hoppings must occur. The Hamiltonian~\eref{Hfermi} describes the
dynamics of a one-dimensional gas of free fermions; the dynamically
active states are those with nonvanishing probability amplitude to
have a fermion-hole pair in neighboring sites. As it commutes with
the fermion-number operator
$\hat{n}_{_{\rm{F}}}\,{=}\,\sum_j\hat{c}^\dagger_j\hat{c}_j$, the
Hamiltonian~\eref{Hfermi} $H_{\rm{R}}$ is reducible: the irreducible
subspaces of ${\cal{H}}_{\rm{R}}$ have an integer eigenvalue
$n_{_{\rm{F}}}$ of $\hat{n}_{_{\rm{F}}}$ and dimension
${{M}\choose{n_{_{\rm{F}}}}}$.


\subsection{The analogy}
\label{ss.analogy}

During an oscillation of the classical NC there are several spheres
at rest and in contact with each other, and some moving spheres. When
a moving sphere hits a sphere at rest, the latter keeps being at rest
and exchanges its momentum with the nearby sphere (upper part of
figure~\ref{analogy}). In the quantum analogue of the NC we are
discussing here, the role of the spheres' momenta is played by the
wave-functions at each lattice site. Rather than the transfer of
mechanical momentum, in the quantum system there is a transmission
along the lattice of a disturbance of the wave-function. This is
represented in the lower part of figure~\ref{analogy}. Furthermore,
in the place of the spheres oscillating at the boundaries of the
chain, in our system we expect to observe the oscillation of the
wave-function amplitude on the lattice ends due to the disturbance
that runs forward and back. The system's wave function at each
lattice site $j$ can be a superposition of the two atomic internal
states $|0\rangle_j$ and $|1\rangle_j$. Under the analogy we propose,
one can for instance associate to the spheres at rest the states
$|0\rangle_j$, and, accordingly, a moving sphere, let us say the
first one, corresponds to a state $a|1\rangle_1+b|0\rangle_1$, i.e.,
a superposition of the two internal states. In terms of atoms this
amounts to consider all sites initially populated by a species-0
atom, but for (a partial superposition with) a species-1 atom in the
first site. This triggers oscillations whose dynamics is ruled by
Hamiltonian~\eref{Hfermi}, and essentially consist in the disturbance
travelling along the lattice, i.e., the solitary species-1 atom
propagates through the chain of species-0 atoms and migrates until
the opposite end, where it is reflected back thus determining the NC
effect. Remarkably, this analogue of the classical propagation is
described in terms of fermions: the most `non-classical' particles.
Note that in both systems the effect of the propagation is perceived
by `observing' the lattice boundaries.

Suppose now that a Bose-Einstein condensate is adiabatically led into
the Tonks-Girardeau regime with one atom per site and with all atoms
in the same internal state, let's say the state
$|0\rangle\equiv\prod_j|0\rangle_j$. The evolved state vector at time
$t>0$ is generally non-separable, and takes the general form
of~\eref{st}. Thanks to the fact that the Hamiltonian~\eref{Hfermi}
is quadratic, one can exactly determine the time evolution for a
generic initial condition. In what follows we are going to consider
different dynamical scenarios that can show up, depending on the
hopping amplitudes $\{\tau_j\}$.

To form an initial state in analogy with that of a classical NC, let
us take that with all atoms in the internal state $|0\rangle_j$, and
give a `kick' only to the first atom,
$|s(0)\rangle=\hat{c}_1^\dagger|0\rangle=|10\cdots0\rangle$,
corresponding to the first atom in the internal state $|1\rangle$ and
all others in the internal state $|0\rangle$. Note that $H_{\rm{R}}$
commutes with the total number operator $\sum_j\hat{n}_j$ and
therefore the state evolves in the one-excitation sector of
${\cal{H}}_{\rm{R}}$.

To calculate the dynamics one has to diagonalize the
Hamiltonian~\eref{Hfermi} in the form
\begin{equation}
 H_{\rm{R}} = \sum_{n=1}^M \omega_n~\hat\eta^\dagger_n\hat\eta_n ~,
\label{Hfermik}
\end{equation}
so the initial state evolves as $|s(t)\rangle=\sum_n
g_{n1}\,\rme^{-\rmi\omega_nt}\hat\eta^\dagger_n|0\rangle$
(see~\ref{a.evol}). The probability amplitude to find a particle in
the $j$-th site at time $t$ is given by
\begin{equation}
 A_j(t)\equiv\langle{0}|\hat{c}_{j}|s(t)\rangle
       = \sum_{n=1}^M g_{n1}g_{nj}~\rme^{-\rmi\omega_nt} ~,
 \label{aj}
\end{equation}
and can be calculated numerically.


\section{Two opposite schemes for a quantum Newton cradle}
\label{s.uniform-perfect}

In this section we examine the possibility of realizing a QNC by
means of the most natural choice of a simply uniform lattice. The
simple theoretical treatment illustrates the mechanism underlying the
cradle's oscillations as well as the reasons why this particular
setup does not behave as expected. This is contrasted with a
different {\em ad-hoc} arrangement of the hopping amplitudes, which,
in spite of being difficult to attain in practice, shows that what is
needed regards the frequency spectrum of the discrete system, namely
that the frequencies be equally spaced.

\subsection{Uniform QNC}
\label{ss.uniform}

\begin{figure}[t]
\hfill\includegraphics[width=130mm,clip=true]{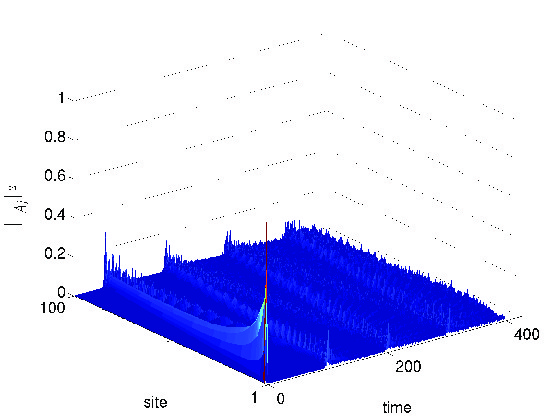}
\caption{Disturbance of the wave-function that travels
forward and backward along the chain.}
\label{figTrappingFlat}
\end{figure}

This is the simplest case, both theoretically and experimentally, and
occurs when all hopping amplitudes are equal, $\tau_j=\tau$, i.e.,
the chain is {\em uniform}. In this case one has~\cite{BanchiV2013}
the orthogonal matrix $g_{nj}=\sqrt{\frac2{M{+}1}}~\sin{k_n}j$ with
$k_n=\frac{\pi n}{M{+}1}$~$(n=1,\dots,M)$ and the eigenvalues
$\omega_n=-2\cos{k_n}$; it follows that the evolving excitation is
described by the sum
\begin{equation}
 A_j(t)=\frac2{M{+}1}\sum_{n=1}^M\sin{k_nj}\sin{k_n}~\rme^{2\rmi t\cos{k_n}}~,
\end{equation} whose square modulus is reported in
figure~\ref{figTrappingFlat}.
It is clearly shown that the initial disturbance of the wave-function
travels along the chain in the form of a wavepacket, which reaches
the opposite end of the chain and is reflected backward. However, one
can clearly see a significant attenuation of the transmitted signal,
an effect essentially due to the destructive interference of the
wave-function components. In other words, after a few bounces, namely
in a time of the order of a multiple of $M$, the initial state
$|1,0\ldots0\rangle$ evolves to a state where the species-1 boson is
delocalized along the chain. This is the generic situation that
occurs in a dispersive system: the wave-function spreads over the
lattice during the time evolution until the initial wavepacket is
lost. A similar phenomenon also occurs in the classical NC if the
masses of the spheres are not identical, i.e., in the non-uniform
case. Evidently, in the quantum analogue, the uniformity of the
system causes dispersion: therefore, it is important to identify
under which conditions such attenuation can be minimized.


\subsection{Perfect QNC}
\label{ss.perfect}

The dynamic decoherence described above can be not only reduced but
even eliminated by letting the tunnelling amplitudes to vary along
the chain with well-defined {\em nonuniform} values. In fact, in the
field of quantum information theory it is known since a few
years~\cite{ChristandlDEL2004,ChristandlDDEKL2005,KarbachS2005,WangSR2011}
that a dispersionless end-to-end quantum-state transmission can be
obtained, for a Hamiltonian like~\eref{Hfermi}, when its
nearest-neighbour couplings are suitably engineered, namely with
proper values of the parameters $\{\tau_j\}$.

\begin{figure}[t]
\hfill\includegraphics[width=130mm,clip=true]{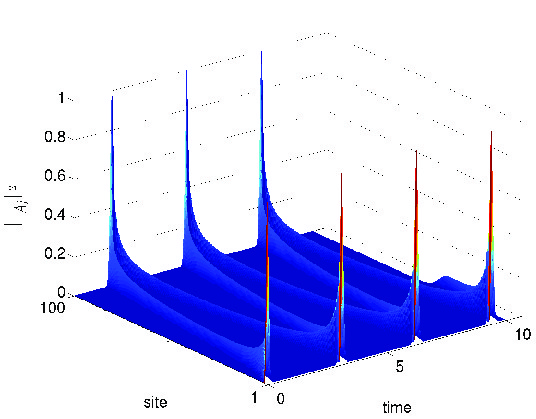}
\caption{Disturbance of the wave-function that travels
forward and backward along the suitably `adapted' chain with
$\tau_j\propto\sqrt{j(M{-}j)}$.}
\label{ajopt}
\end{figure}

Looking at~\eref{aj}, it is indeed apparent that if all frequencies
are `linear', namely if $\omega_n\,{=}\,\Omega_0\,{+}\,\Omega\,n$,
then, setting the half-period time $t'\,{=}\,\pi/\Omega$, one finds
that
$A_j(2t')\,{=}\,\rme^{-2\rmi\Omega_0t'}\,A_j(0)\,{=}\,\rme^{-2\rmi\Omega_0t'}\,\delta_{j1}$,
i.e., the initial state is exactly reproduced (up to an overall
phase) and the dynamics has a period $2 t^\prime$. Furthermore, for a
mirror-symmetric chain ($\tau_j=\tau_{M-j}$) one has alternatively
mirror-symmetric/antisymmetric eigenvectors~\cite{CantoniB1976},
$g_{n,M{+}1{-}j}=(-)^ng_{nj}$, and therefore
\begin{equation}
 A_j (t^\prime) = \rme^{-\rmi\Omega_0 t^\prime}\,\delta_{jM} ~,
\end{equation}
meaning that at the time $t^\prime$ the initial state is perfectly
mirrored and the initial excitation $\hat{c}_1^\dagger|0\rangle$ is
fully transferred to the opposite end of the chain
$\big|\,s(t^\prime)\big\rangle= \rme^{-\rmi\Omega_0
t^\prime}\,\hat{c}_M^\dagger|0\rangle$. However, in order to obtain
eigenvalues $\{\omega_n\}$ suitable for perfect transfer, the
couplings $\{\tau_j\}$ must be individually
engineered~\cite{ChristandlDEL2004}. This can be rapidly proven by
considering a spin-$S$ subjected to a field along the $x$-direction,
i.e., whose Hamiltonian is
$H=\Omega\hat{S}^x=\frac12\Omega(\hat{S}^+\,{+}\,\hat{S}^-)$, with
\begin{equation}
 S^+=\sum_{m=-S}^S \sqrt{(S{+}1{+}m)(S{-}m)}~|m{+}1\rangle\langle{m}|~;
\end{equation}
replacing $2S{+}1=M$, $j=S{+}1-m$, and identifying
$\hat{c}^\dagger_j|0\rangle=|S{+}1{-}j\rangle$, one exactly maps the
uniform rotation $\rme^{-\rmi t\Omega\hat{S}^x}$ onto the perfect
excitation transmission forth and back along the array with couplings
$\tau_j=\Omega\sqrt{j(M{-}j)}$, which indeed yield equally spaced
eigenvalues $\omega_n=\Omega(n{-}\frac{M{-}1}2)$. So, in principle,
there is a way to obtain a perfect quantum cradle, whose behaviour is
illustrated in figure~\ref{ajopt}. Nevertheless, one has to recognize
that its experimental realization by means of the accurate tuning of
each tunnel coupling is an apparently hopeless task.
It is worthwhile to mention that, in a different context, a cold-atom
system undergoing the dynamics of a large rotating spin has been
realized in~\cite{CatAtomChip}.


\section{Two realistic schemes for a quasi-perfect QNC}
\label{s.qperfect}

On the basis of the discussion in the preceding section, we are going
here to show that it is possible to minimally modify the least
demanding uniform lattice in order to strongly improve the cradle's
performance. The main observation is that one can limit the requisite
of a linear spectrum only to those modes which are actually activated
by triggering a perturbation at an endpoint of the chain.

\subsection{Quasi-uniform array}
\label{ss.qperfect1}

There exists a simpler way for the actual realization of a
high-quality QNC in an essentially uniform chain, such that the need
of engineering is littlest. A natural strategy, in order to move
towards the shape of the hopping amplitudes of Sec.~\ref{ss.perfect},
is that of weakening the extremal $\tau_j$'s, for instance by acting
with a transverse beam to increase the potential barriers between the
endpoint wells of the optical lattice. Indeed, keeping the
requirement of a mirror-symmetric chain~\cite{Kay2010}, one can
minimally modify a uniform chain taking equal couplings,
$\tau_j=\tau$, but for the ones at the edges,
$\tau_1=\tau_{M{-}1}=x\,\tau$, with $x\,{<}\,1$, and look for the
best transfer conditions. In~\cite{BanchiV2013} it is shown that the
eigenvalues can still be written as $\omega_n=-2\cos{k_n}$, but the
pseudo-wavevectors
\begin{equation}
 k_n = \frac{\pi n+2\varphi_{k_n}}{M{+}1}~~~~ (n=1,\,..,\,M)~,
\end{equation}
varying in the interval $(0,\pi)$ are no more equally spaced, being
affected by the {\em shifts}
\begin{equation}
 \varphi_k = k -
 \cot^{-1}\big(\textstyle\frac{x^2}{2{-}x^2}\cot{k}\big) \, .
\end{equation}
The discussion made above about the ideal case shows that a `linear'
dispersion relation guarantees coherent transmission of the initial
excitation. However, the eigenvalues $\omega_n\,{=}\,-2\cos{k_n}$ of
the quasi-uniform chain are almost linear just in a neighborhood of
the inflection point, $k_n\sim\pi/2$. Looking more closely at the
expression of the amplitude~\eref{aj} at the last site, that, using
the mirror-symmetry property $g_{n,M{+}1{-}j}=(-)^ng_{nj}$, reads
\begin{equation}
 A_M(t) =  \sum_{n=1}^M  g_{n1}^2 \rme^{\rmi(\pi-\omega_{k_n} t)} \, ,
\label{aMgen}
\end{equation}
one can suppose that a good deal would be to have the prefactor
$g_{n1}^2$, which can be thought of as a normalized probability,
strongly peaked just in the linear zone where $n\sim(M{+}1)/2$, in
such a way that the dynamics involves coherent modes. We have seen
that this is not the case for the fully uniform chain, because
$g_{n1}^2\sim\sin^2k_n$ has its maximum in the desired zone but has
too broad a shape. It is shown in~\cite{BACVV2011} that the shape of
$g_{n1}^2$ shrinks when $x$ decreases. As expected, this effect
improves (the absolute value of) the transmission
amplitude~\eref{aMgen} up to a maximum that depends on $M$: even for
an infinitely long chain this optimal value is still finite and
larger than $0.853$, the optimal coupling being
$x\simeq1.03\,M^{-1/6}$. When $x$ gets smaller the distribution
$g_{n1}^2$ gets narrower and narrower, however what prevents from
getting perfect transfer is the obvious fact that by varying $x$ one
also affects the shape of the dispersion relation $\omega_{k_n}$,
perturbing its `quasi-linear' behavior. It is natural to observe
that, in order to have an almost independent control over these {\em
two} effects (shrinking of the weighting density $g_{n1}^2$ and
deformation of $\omega_{k_n}$) {\em two} parameters are needed. As a
matter of fact taking into play also the second bonds
$\tau_2=\tau_{M{-}2}=y\tau$ makes the difference and allows one to
guarantee a response larger than $0.987$ when the coupling are tuned
as $x\simeq2M^{-1/3}$ and $y\simeq2^{3/4}M^{-1/6}$~\cite{BACVV2012}.
Moreover, optimal response is obtained also in a wide neighborhood of
the optimal couplings, so there is no need to fine-tune them.


\subsection{Uniform array with a Gaussian trap}
\label{ss.qperfect2}

\begin{figure}[t]
\hfill\includegraphics[width=130mm,clip=true]{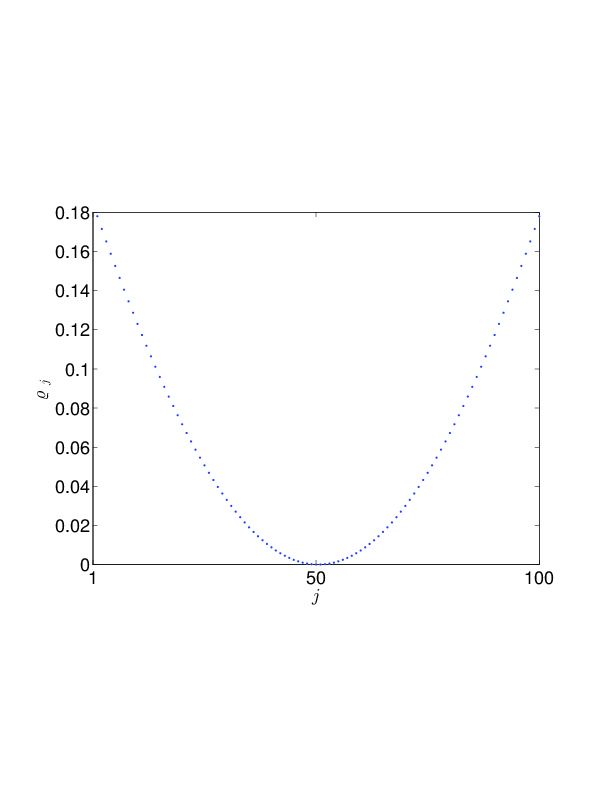}
\caption{The trapping potential generates a site-dependent energy-offset
$\varrho_j$, for $j=1,\dots,M$. The figure shows $\varrho_j$ as a
function of the site index $j=1,\dots,M$.}
\label{TrapPot}
\end{figure}

The last scheme we propose, considers a configuration that can be
better realized in an experiment. Besides the uniform one-dimensional
optical potential, we add a trapping potential that generates a
site-dependent energy-offset $\varrho_j$, for $j=1,\dots,M$, with a
Gaussian profile (figure~\ref{TrapPot}). Furthermore, we choose the
initial state $|s(0)\rangle = \sum_jz_j\hat{c}^\dagger_j|0\rangle$,
that represents a Gaussian wave-packet along the lattice. The site
wave-function in fact is $z_j=A\exp(-(j-x_0)^2/\varsigma^2)$, where
$A$ is a normalization factor and $x_0$ and $\varsigma$ are the
centre and the width of the packet, respectively.

\begin{figure}[t]
\hfill\includegraphics[width=130mm,clip=true]{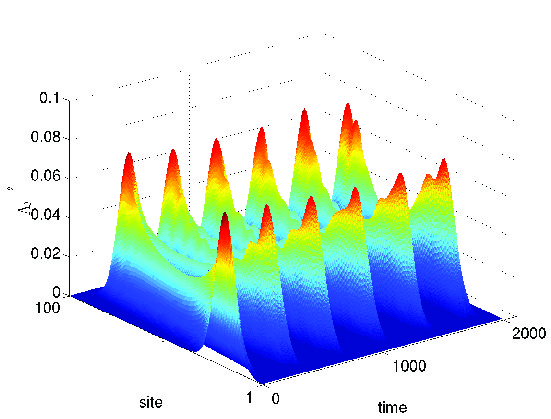}
\caption{Disturbance of the wave-function that travels forward and
backward in the uniform chain, $\{\tau_j = 1 \}$, with a superimposed
Gaussian confining potential. When the wave-packet moves toward a
lattice's end, its speed is slowed down by the trapping potential,
until the motion is inverted. Then the packet is accelerated in the
opposite direction, as long as, it reaches the centre of the
potential.}
\label{TrappingGaussian}
\end{figure}

Such a setup is a more realistic configuration with respect to the
previous ones. In fact, in the  schemes that we have illustrated so
far, the bounce of the disturbance of the wave-function at the
lattice ends, takes place because of the open-boundary conditions. On
the contrary, in the present setup,  the wave-packet oscillates
inside the trapping potential and its speed inversion is caused by
the forces generated by the trapping potential. In
figure~\ref{TrappingGaussian} it is evident that the packet never
reaches the lattice ends. In fact, when the wave-packet moves from
the trapping centre towards a lattice's end, its speed is slowed down
by the trapping potential, until the motion is inverted and the
packet is accelerated in the opposite direction.

\begin{figure}[t]
\hfill\includegraphics[width=130mm,clip=true]{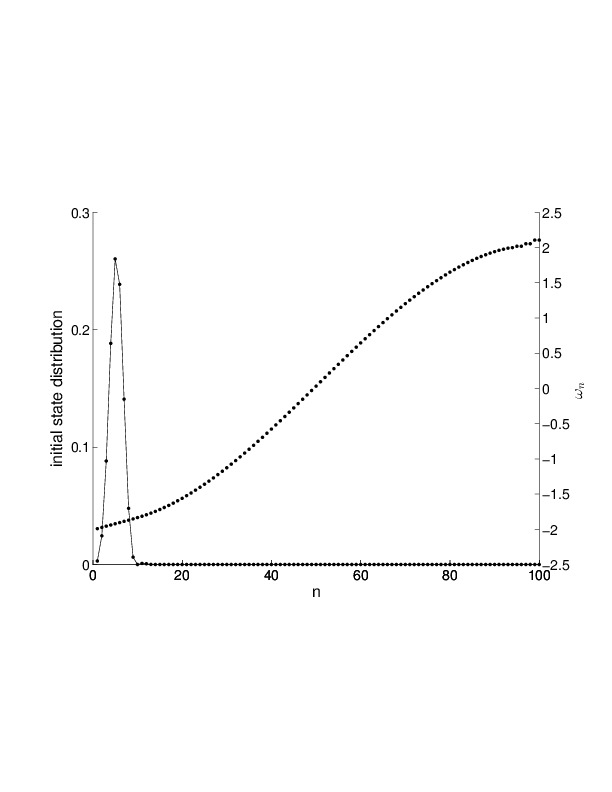}
\caption{Eigenvalues $\omega_n$ in presence of the Gaussian trapping
potential (right $y$-axis) and mode components of the initial state
$|s(0)\rangle=\sum_jz_j\hat{c}^\dagger_j|0\rangle$ (left $y$-axis),
showing that this state only involves a linear portion of the
`dispersion relation'.}
\label{IS-EL}
\end{figure}

Figure~\ref{TrappingGaussian} illustrates the result of a simulation
made with an optical lattice of $M\,{=}\,100$ sites and an initial
state with $x_0\,{=}\,20$ and $\varsigma\,{=}\,10$. The on-site
energy offset is $\varrho_j\,{=}\,\exp(-(j-x_m)^2/\vartheta^2)$, with
$x_m\,{=}\,50$ and $\vartheta\,{\approx}\,110$. The `dispersion
relation' $\omega_n$ shown in figure~\ref{IS-EL}, displays intervals
with approximately equal spacing, which is the condition leading to
quasi-perfect transmission. To this goal, the initial state is taken
as a narrow superposition of eigenmodes chosen in an almost linear
region as shown in figure~\ref{IS-EL}. Figure~\ref{TrappingGaussian}
proves indeed that the system displays a high transmission amplitude.

\section{Conclusions}
\label{s.conclusions}

We have investigated an experimental framework that could realize a
quantum analogue of Newton's cradle, starting from a Bose-Einstein
condensate of two atomic species (i.e., atoms with two internal
states) trapped in a one-dimensional tube with a longitudinal optical
lattice; the system is kept in a strong Tonks-Girardeau regime with
maximal filling factor, so that the lattice sites contain one atom
and at each site the wave function is a superposition of the two
internal atomic states. We have shown that the tunnelling between
sites makes the system equivalent to a free-fermion gas on a finite
lattice. In these conditions, one can trigger a disturbance of the
wave function at one lattice end and this perturbation starts
bouncing back and forth from the ends, just as the extremal spheres
in the classical Newton's cradle: the analogy associates the
propagation of a wave-function disturbance with the transmission of
mechanical momentum.

However, in the quantum system the travelling wave undergoes
decoherence, a phenomenon that makes a uniform lattice (i.e., with
uniform tunnelling amplitudes) almost useless; still, it is
known~\cite{ChristandlDEL2004} that a suitable arrangement of the
tunnelling amplitudes can even lead to a virtually perpetual cyclic
bouncing. These possible schemes illustrate two opposite situations:
the first is easy to realize but gives a scarce result, the latter
would be perfect but requires an almost impossible engineering.

That's why we looked for compromises that, minimizing the required
experimental adaptation of the interactions, give almost perfect
quantum Newton's cradles. In this respect, we proposed two schemes:
the first one involves a weakening of the extremal pairs of couplings
and can lead to an amplitude response of 99\%, i.e., the dynamics
could be observed for tenths of oscillations; in the second we
considered a longitudinal trapping potential with a Gaussian shape.
Of course, the possibility to obtain quantum systems that allow
high-quality wave transmission is not only relevant from the
speculative point of view, but also in the field of the realization
of quantum devices like atomic interferometers, quantum memories, and
quantum channels. Nevertheless, realizing the quantum Newton's cradle
we proposed would be stirring by itself for the insight it would give
into the entangled beauty of Quantum Mechanics.


\ack
R.~F. thanks Vittorio Penna and Pierfrancesco Buonsante for their
help in the present investigation.


\appendix
\section{Bose-Hubbard model}
\label{a.BHmodel}

We start with a mixture of bosonic atoms with two different internal
states, subjected to a strong transverse trapping potential with
frequency $\omega_\perp \gg \mu/\hbar$, where $\mu$ is the chemical
potential. The quantum dynamics of an ultracold dilute mixture of
bosonic atoms with two internal states is described by the
Hamiltonian operator
\begin{eqnarray}
 {\hat {H}} = \!\sum^2_{\alpha=1} \!  \int \rmd{\bm{r}}
 \hat\psi^\dagger_\alpha(\bm{r})
 \left [ -\frac{\hbar^2}{2m}\nabla^2
 + V^\alpha_{ext} (\bm{r}) \right ] \hat\psi_\alpha(\bm{r}) +
\nonumber\\
 \frac{\pi \hbar^2 }{m} \! \sum^2_{\alpha, \beta = 1}  a_{\alpha \beta}
 \int \! \rmd\bm{r}
 \hat\psi^\dagger_\alpha(\bm{r})\hat\psi^\dagger_\beta(\bm{r})
 \hat\psi_\beta(\bm{r})\hat\psi_\alpha(\bm{r}) \, ,
\label{mbh}
\end{eqnarray}
where $\alpha$ and $\beta$ take the values $1,2$, corresponding to
the internal states, the boson-field operator
$\hat\psi_\alpha(\bm{r})$ ($\hat\psi_\alpha^\dagger(\bm{r})$)
annihilates (creates) an atom at $\bm{r}=(r_x,r_y,r_z)$ in the
internal state $\alpha$. $V^\alpha_{ext}$ is the external trapping
potential seen by the atoms in the state $\alpha$. The nonlinear
interaction term depends on the intraspecies and interspecies
scattering lengths $a_{\alpha \beta}$, and on the atomic mass $m$. We
consider the case of repulsive atomic interactions, thus
$a_{\alpha\beta}>0$ for $\alpha,\beta=1,2$. The external trapping
potential of the optical lattice reads
$V^\alpha_{ext}(\bm{r})=V^\alpha\sin^2(k_Lr_x)+m\omega_\perp^2(r^2_y+r^2_z)/2$,
where $k_L$ is the laser  mode which traps the atoms.
Accordingly, the physics is effectively one-dimensional.
Following~\cite{Jaksch98,Jaksch05}, \eref{mbh} is transformed into an
effective quantum (Bose-Hubbard) Hamiltonian that describes, within the second
quantization formalism, the boson mixture dynamics in an optical
lattice.
The boson-field operator can be rewritten in terms of the Wannier functions
${u}^\alpha_{j}(\bm{r})$ as (see~\cite{Jaksch98})
\begin{equation}
\hat\psi_\alpha(\bm{r},t) = \sum^{M}_{ j=1}
{u}^\alpha_{j}(\bm{r}) \hat a_{\alpha{j}}(t)
\, ,
\label{fieldoperator}
\end{equation}
where the boson operator $\hat a_{\alpha{j}}$ ($\hat
{a}_{\alpha{j}}^\dagger$) destroys (creates) an atom in the internal
state $\alpha$ at the lattice site $j$. By
substituting~\eref{fieldoperator} into Hamiltonian~\eref{mbh}, and
keeping the lowest order in the overlap between the single-well wave
functions, one obtains the effective one-dimensional Bose-Hubbard
Hamiltonian~\eref{BHH}. Where, $U$ and $U_1, U_2$ are the inter- and
intra-species onsite interaction energy, respectively, and $t_\alpha$
is the hopping amplitude of the species $\alpha$. These constants can
be estimated, e.g., in the tight-binding approximation one obtains
$U_\alpha ={2\pi \hbar^2 a_{\alpha \alpha} }/{m} \int d{x}
|{u}^\alpha_{ j}|^4 $ and $t_\alpha = -2 \int d{x}
\bar{u}^\alpha_{ j} [\frac{\hbar^2}{2m}\nabla^2 + V^\alpha_{ext}
]u^\alpha_{{ j+1} }$~\cite{Jaksch98}.


\section{Fermionization}
\label{a.fermioni}

Considering the limit of strong on-site repulsion (large $U$'s) we
can separate from the Hamiltonian~\eref{BHH} a perturbative term $V$,
writing $H=H_0\,{+}\,V$. Thus, up to an irrelevant constant factor it
results
\begin{eqnarray}
 H_0 &=& \sum_j \big[ U_0\hat{n}_{0j}(\hat{n}_{0j}{-}1)
                   +  U_1\hat{n}_{1j}(\hat{n}_{1j}{-}1)
\nonumber\\
 &&~~~~~~~ + \frac{U}{2}(2\hat{n}_{0j}\hat{n}_{1j}
 +1{-}\hat{n}_{0j}{-}\hat{n}_{1j}) \big]
\\
 V &=&\!\!\! -\!\!\! \sum_{j=1}^{M-1} \!\! \big[
             t_{0j}\, \hat{a}_{0j}^\dagger\hat{a}_{0,j+1}
             +t_{1j}\, \hat{a}_{1j}^\dagger\hat{a}_{1,j+1}
             +\rm{h.c.}\big]~.
\label{e.V}
\end{eqnarray}
The recipe for the effective Hamiltonian restricted to the Hilbert
space ${\cal{H}}_{\rm{R}}$ of the states for which the expectation
value of $H_0$ vanishes (i.e., $\hat{n}_{0j}{+}\hat{n}_{1j}\,{=}\,1$)
is~\cite{prlKuklovSvistunov,prlDuanDemlerLukin,praPachosRico}
\begin{equation}
 \langle\alpha|H_{\rm{R}}|\beta\rangle =
  - \sum_{\psi} \frac{\langle\alpha|V|\psi\rangle\langle\psi|V|\beta\rangle}
                {\langle\psi|H_0|\psi\rangle}~,
\end{equation}
where the virtual states $\psi$ are necessarily outside
${\cal{H}}_{\rm{R}}$. In order that
$V|\psi\rangle\in{\cal{H}}_{\rm{R}}$ both matrix elements in $V$ must
involve the same site pair $(j,j{+}1)$; for each pair, out of 16
possible terms, 6 do not vanish and it turns out that the effective
interaction can be eventually written as
\begin{eqnarray}
 H_{\rm{R}} &=&
 \sum_{j=1}^{M-1} \big[
   - \tau_j\,(\hat b^\dagger_{j} \hat b_{j+1}+\hat b^\dagger_{j+1} \hat b_{j})
   +\gamma_j\,\hat b^\dagger_j \hat b_j \hat b^\dagger_{j+1} \hat b_{j+1}
\nonumber\\
   &&\hspace{8mm}
   +\sigma_j\, (\hat b^\dagger_j \hat b_j + \hat b^\dagger_{j+1} \hat b_{j+1})
   \big] ~,
\end{eqnarray}
where a further additive constant is neglected and
\begin{eqnarray}
 \tau_j &=& 2 \frac{t_{0j}t_{1j}}{U}
~,~~
 \gamma_j = 2 \left[ \frac{t_{0j}^2{+}t_{1j}^2}{U}-\frac{t_{0j}^2}{U_0}
 -\frac{t_{1j}^2}{U_1} \right]
\\
 \sigma_j &=& 2 \frac{t_{0j}^2}{U_0}-\frac{t_{0j}^2{+}t_{1j}^2}{U}~;
\end{eqnarray}
$\hat{b}_j=\hat{a}_{0j}^\dagger\hat{a}_{1j}$ and $\hat{b}_j^\dagger$
are operators in ${\cal{H}}_{\rm{R}}$ such that
$\{\hat{b}_j,\hat{b}_j^\dagger\}=1$, while between different sites
all commutators vanish. In the one-dimensional case they can be
converted into Fermion operators by the transformation~\eref{e.JW};
if the interactions do not depend on the atomic internal state
$H_{\rm{R}}$ turns into~\eref{Hfermi}.


\section{Time evolution}
\label{a.evol}

The exact eigenstates of Hamiltonian~\eref{Hfermi} are derived by
means of a linear transformation, $\hat\eta_n\,{=}\,g_{nj}\hat{c}_j$
and $\hat\eta^\dagger_n\,{=}\,g^*_{nj}\hat{c}^\dagger_j$. Requiring
it to be canonical, i.e., imposing
$\{\hat\eta_n,\hat\eta^\dagger_{n'}\}=\delta_{nn'}$, one finds that
the $M{\times}M$ matrix $g_{nj}$ must be unitary,
$g^\dagger{g}=gg^\dagger=\mathbb{1}$. Actually, $g$ is also real
(hence, orthogonal), as it diagonalizes the real symmetric matrix
$A_{jj'}=(\tau_j\delta_{j+1,j'}{+}\,\tau_{j'}\delta_{j-1,j'})$, i.e.,
$\sum_{jj'}g_{nj}A_{jj'}g_{n'j'}=\omega_n\,\delta_{nn'}$; the
diagonalized Hamiltonian has the form~\eref{Hfermik}. When most of
the couplings $\tau_j$ are equal, a convenient general method for
solving the eigenvalue problem can be found in~\cite{BanchiV2013}.


\end{document}